\documentclass[preprint,12pt]{elsarticle}

\usepackage{amssymb}
\usepackage{xcolor}
\usepackage{overpic}
\usepackage{subfig}
\usepackage[margin=1in]{geometry}
\usepackage{mathtools}

\usepackage{amsmath}

\DeclareMathOperator{\ODESolve}{ODESolve}

\usepackage{changes}

\journal{XXX}

\begin{document}

\begin{frontmatter}



\title{First Contact: Data-driven Friction-Stir Process Control }

\author{James Koch\corref{cor1}, Ethan King, WoongJo Choi, Megan Ebers, David Garcia, Ken Ross, Keerti Kappagantula}

\cortext[cor1]{Pacific Northwest National Laboratory, 902 Battelle Blvd., Richland, WA, 99354}

\begin{abstract}

This study validates the use of Neural Lumped Parameter Differential Equations for open-loop setpoint control of the plunge sequence in Friction Stir Processing (FSP). The approach integrates a data-driven framework with classical heat transfer techniques to predict tool temperatures, informing control strategies. By utilizing a trained Neural Lumped Parameter Differential Equation model, we translate theoretical predictions into practical set-point control, facilitating rapid attainment of desired tool temperatures and ensuring consistent thermomechanical states during FSP. This study covers the design, implementation, and experimental validation of our control approach, establishing a foundation for efficient, adaptive FSP operations.

\end{abstract}

\end{frontmatter}

\section{Introduction} \label{sec:intro}

Friction-stir processing (FSP) is a robust method for modifying materials to achieve desired properties because of its unique ability to achieve microstructural characteristics of interest \cite{mishra2005friction, heidarzadeh2021friction}. FSP uses a non-consumable rotating tool that locally heats and plastically deforms materials through the act of material mixing. However, process control of FSP remains a challenge: defects can form (\textit{e.g.}, tunneling defects, ``kissing bonds'', etc.) that are not desirable in the final product as they reduce sample performance \cite{kah2015investigation}. Setting appropriate thermomechanical conditions at the beginning of the process is critical for maintaining sample integrity and minimizing the risk of developing these defects. 

In our recent work, we proposed an innovative approach to modeling temperature time series during the plunge and dwelling stages of FSP on 316L stainless steel \cite{koch2024neural}. Leveraging a data-driven methodology integrated with classical engineering heat transfer techniques, we demonstrated the potential of the Universal Differential Equation (UDE) modeling paradigm \cite{rackauckas2020universal} to construct interpretable models capable of predicting the thermal profiles observed in experiments and enabling control strategies. In this study, we extend these findings by implementing and validating our control strategies on actual hardware. Our aim is to transition from exclusively model temperature predictions to practical set-point control for the FSP plunge sequence, thereby ensuring consistent thermomechanical conditions during the tool travel over the sample during FSP. 

This work details our methodology, experimental implementation, and testing of our control system on hardware setups. Section \ref{sec:methods} provides an overview of both the system identification and model based control tasks. Section \ref{sec:results} summarizes the experimental campaign and performance metrics for this effort.

\begin{figure*}[h!]
\centering
        \begin{overpic}[width=6.5in]{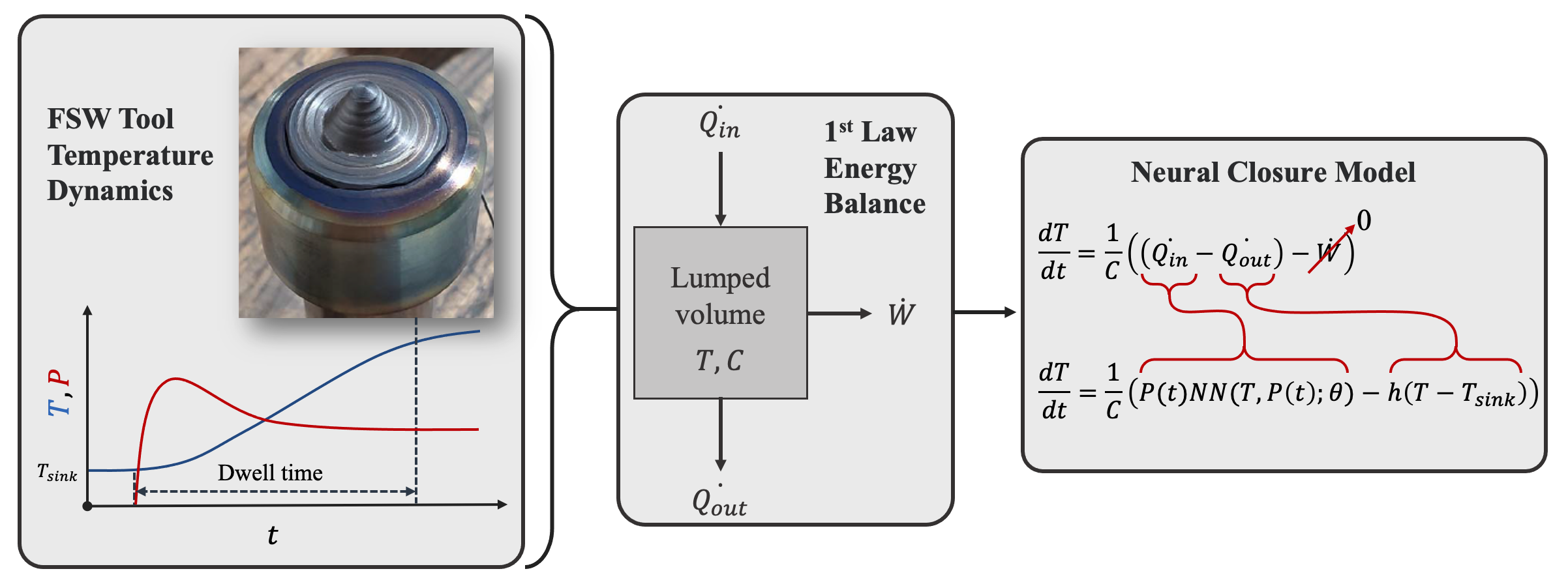}  
        \put(2,33){(a)}
        \put(40,28){(b)}
        \put(65.75,25.5){(c)}
        \end{overpic}  
        \caption{\textit{Friction-stir processing} (FSP) uses a rotating tool to heat and plastically deform material of a workpiece. In (a), shown are notional time histories for power and tool temperature during the plunge sequence of FSP. After reaching a desired temperature set point, the plunge sequence concludes and the tool traverse in the sample commences. Here, we seek to quickly obtain a set point temperature through data-driven open-loop control to reduce material wastage. In (b), we construct a first-principles physics model of the system, with the associated differential equation in (c). While the system-level physics may be adequately described by the First Law of Thermodynamics, the relationships between control inputs and the system dynamics may not be known. We insert a neural network in (c) to approximate the unknown physics present in the system.}
        \label{fig:overview}
\end{figure*}

\section{Methodology} \label{sec:methods}

\subsection{Neural Lumped-Parameter Differential Equations}

During FSP, the tool rotates in the workpiece and owing to the friction and mechanical work, the temperature of the processed region is increased. The conversion of machine input power to internal heating of the workpiece is governed by the First Law of Thermodynamics, \textit{i.e.} all input/output energy must be accounted for during processing. This is true at any observational scale, whether that be at the molecular level or system-wide. Prior to the tool traverse along the sample in a predetermined path, it has to be plunged into the workpiece and process conditions have to be constantly evolved to achieve the desired control parameters such as a constant tool temperature. In the setting of the plunge and dwell stages, we choose to construct a model at the granularity of the observed temperature and machine processing parameters using an energy accounting scheme derived from the First Law of Thermodynamics, as shown in Fig. \ref{fig:overview} and first presented in Koch et al. \cite{koch2024neural}.

A time-dependent energy accounting scheme can be framed as a first-order ordinary differential equation (ODE) for energy: $dE/dt= \dot{Q}- \dot{W} $, where $E$ is energy, $\dot{Q}$ is the rate of heat input into the system, and $\dot{W}$ is the work done by the system on its surroundings. Given that we are interested in how power is converted into a temperature (with no interactions with the surroundings), this can be re-written in a more useful form that relates temperature dynamics with input and output heating terms for a lumped volume with thermal capacitance $C$;
\begin{equation} \label{eq:system}
    \frac{dT}{dt} = \frac{1}{C}\left(\dot{Q}_{in}- \dot{Q}_{out}\right) = \frac{1}{C} \left(\eta P\left(t\right)-h\left(T-T_{sink} \right)\right) \coloneqq f\left(T,P\left(t\right);\theta\right),
\end{equation}
where we define $f$ as a stand-in for the right-hand-side (RHS) of the ODE. Further, we have assumed that temperature dynamics are directly related to the difference in heating through a power input $P(t)$ (scaled by some unknown conversion factor $\eta$) and linear heat transfer out of the system. While we have neglected many physical effects in this minimum viable model for temperature dynamics, the essential physics that govern our observable are present, and the overall behavior can be directly linked back to first-principles physics. In this model, the conversion factor $\eta$ from power to internal heating of the workpiece is unknown, as is the capacitance of the lumped volume $C$ and the linear heat transfer coefficient $h$. 

The data science task is to tune these unknowns such that the model best reproduces experimental data. Recent advances in machine learning (ML) for dynamical systems has enabled gradient-based optimization for systems that obey differential equation relationships. Termed Neural ODEs \cite{chen2018neural}, this class of methods allows for automatically differentiating ODE-based model output with respect to inputs and parameters, enabling deep learning integration with physical models. Here, we replace the conversion factor $\eta$ with a neural network $\eta(T,P;\theta):\mathbb{R}^2\rightarrow(0,1)$  that relates tool temperature and input power to a conversion efficiency. For this work, the neural network is a simple feed-forward multi-layer perceptron with a single hidden layer of size 15 and a sigmoid activation function. The output layer is similarly sigmoid-activated to constrain the output to be in the range (0,1). Fitting this model to available data constitutes the \textit{system identification} (system ID) problem, formalized in Section \ref{sec:system_id}.

Given a differentiable system ID model, model weights can be frozen and a trainable network inserted as a stand-in for exogenous inputs. In this case, the exogenous input is the controllable FSP machine power profile, which can be tuned to guide the tool temperature to a desired set temperature or temperature-time profile. This constitutes the open-loop control task, formalized in Section \ref{sec:control}. Note that while this work is restricted to open-loop control, this framework can be readily extended to real-time feedback control \cite{drgovna2022differentiable} provided there is experimental infrastructure to support real-time sampling and control.

\subsection{System Identification Task} \label{sec:system_id}
The system identification task is to minimize the difference between the system dynamics model of Eq. \ref{eq:system} and experimentally obtained data. Given a vector of model parameters $\theta$, this can be written as finding those that minimize the model-data temperature mismatch:
\begin{equation} \label{eq:closure}
\begin{split}
\underset{\theta}{\text{arg~min}} &~~ \sum_{i=1}^{I} \sum_{k=1}^{K_i}  ||T^{(k)}_i - \hat{T}^{(k)}_i||_2^2 \\
   ~  \text{s.t.~~} T_i^{(k+1)}\ & = \ODESolve\left(f\left(T,P\left(T\right);\theta\right),T_i^{(k)},\left(t,t + \Delta t\right)\right), \\
   ~  T_i^{\left(0\right)} &= \hat{T}_i^{\left(0\right)} ,
\end{split}
\end{equation}
where $i \in I$ denotes the index of each experimental run, $k \in K_i$ is the number of data points in each $i$-th experimental data time series, and the system dynamics model of Eq. \ref{eq:system} is integrated through a standard numerical integrator, denoted by the $\ODESolve$. Here, the $\hat{\cdot}$ denotes membership of the dataset. 

The optimization problem is solved in the Julia computing ecosystem. We use DifferentialEquations.jl and DiffEqFlux.jl \cite{rackauckas2017differentialequations} to formulate the problem. Our integrator is a standard 4-th order adaptive Runge-Kutta integrator. Optimization is through an scheme with both the Adam and BFGS optimizers; we begin with 200 epochs of the Adam optimizer at a learning rate of 0.01 followed by BFGS until convergence (as flagged by detection of a local minimum). Training is performed on a mid-grade consumer laptop. For the limited amount of data available for model training, wall-clock time is negligible; on the order of 5 minutes. 

\subsection{Open-Loop Control Task} \label{sec:control}
Given a successful system ID task, we have access to a set of parameters $\theta$ that, when inserted into the system dynamics model of Eq. \ref{eq:system}, best reproduce observed behaviors subject to the physical priors included in the model. The open-loop control task is one of \textit{optimizing} the system ID model inputs to affect a change in system dynamics. For such tasks, we ``freeze'' our tuned parameter set $\theta$ and now include a new set $\phi$ that parameterizes control inputs. For closed loop control, a controller has access to system observations to adapt control inputs in real time. In open-loop control, there is no feedback, as in the case for the present study. We focus on \textit{trajectory optimization}, where we wish to craft an open-loop control signal to guide our system's state to a desired end point. 

We prescribe a multi-objective loss that we seek to minimize in our trajectory optimization. Losses include a setpoint loss, an energy consumption penalty, a smoothing penalty, and an overshoot penalty. Our problem formulation follows that of Section \ref{sec:system_id} albeit with $\theta$ already obtained and fixed:

\begin{equation} \label{eq:control}
\begin{split}
\underset{\phi}{\text{arg~min}} &~~ \mathcal{L}_{\text{setpoint}} + \mathcal{L}_{\text{energy}} + \mathcal{L}_{\text{smoothing}} + \mathcal{L}_{\text{overshoot}}  \\
   ~  \text{s.t.~~} T^{(k+1)}\ & = \ODESolve\left(f\left(T,P\left(t;\phi\right);\theta\right),T^{(k)},\left(t,t + \Delta t\right)\right), \\
   ~  T^{\left(0\right)} &= \hat{T}^{\left(0\right)} , \\
   ~  \mathcal{L}_{\text{setpoint}} &= \frac{\lambda_1}{K} \sum_{k=1}^{K}  \left(T^{(k)} - T_{\text{s.p.}}\right)^2, \\
   ~  \mathcal{L}_{\text{energy}} &= \lambda_2 \int_0^{t_{\text{end}}} P\left(t;\phi\right) dt, \\
   ~  \mathcal{L}_{\text{smoothing}} &= \frac{\lambda_3}{K-1} \sum_{k=1}^{K-1} P\left(t^{(k+1)};\phi\right) - P\left(t^{(k)};\phi\right) , \\  
   ~  \mathcal{L}_{\text{overshoot}} &= \lambda_4 \sum_{k=1}^{K}  \text{relu}\left(T^{(k)} - T_{\text{s.p.}}\right), \\  
\end{split}
\end{equation}

The optimization objective is comprised of several terms that are crafted to promote certain system behaviors. The setpoint loss promotes reaching the desired end temperature after a model rollout (\textit{e.g.}, after 120 seconds). The energy penalty integrates the control action over the duration of the simulation and seeks to force this value to zero. The smoothing penalty acts to make the control profile smooth in time such that mechanically infeasible or unsafe operations are discouraged in the optimization of the power profile. Lastly, the overshoot term penalizes the accumulation of all model evaluations above the setpoint temperature.

Note that in Eq. \ref{eq:control}, we make no assumptions on the structure of the parametric power input, $P(t;\phi)$. This methodology requires that the function mapping time to power is differentiable, such as a neural network, polynomial, or other analytic mapping, such that sensitivities can be calculated during gradient-based optimization.

\subsection{FSP Experimentation}
In practice, the plunge sequence of FSP is dictated by reference tracking for a desired power profile curve. The FSP machine attempts to reach a desired power level by adjusting spindle speed and torque via a tuned Proportional-Integral-Derivative (PID) controller. The built-in machine control software expects the user to provide a third-order polynomial for the plunge sequence power profile. For this set of experiments, a trained neural lumped parameter system ID model is optimized via the procedure in Section \ref{sec:control} to provide this third-order polynomial power profile that can be directly input into the machine's built-in power control software. Thus, we define $P$ as:
\begin{equation} \label{eq:power}
    P\left(t;\phi \right) = \phi_1 + \phi_2 t + \phi_3 t^2 + \phi_4 t^3.
\end{equation}
The idle power draw of the FSP machine (i.e. the power draw prior to tool-workpiece contact) is approximately 1 kW. Therefore, to promote continuity pre- and post-contact, we enforce that $\phi_1 = 1 \text{kW}$ for all experiments.

Once the FSP machine flags the tool as having reached the desired set tool temperature, the machine performs an automated ``handoff" between power control to temperature setpoint control, the plunge sequence is concluded, and the tool traverse along the preset path in the sample begins. 

\section{Results} \label{sec:results}

Two batches of system ID and control tasks were performed. The first batch consisted of fitting Eq. \ref{eq:system} to seven experiments performed in an initial effort to compile data for ML. Using this data and trained model, we estimated optimal polynomial power profiles to reach two unseen set point temperatures; 775 and 750 degrees Celsius. Results from deploying these power profiles on hardware are presented in Section \ref{sec:batch_1}.

After performing the first batch of validation experiments, we incorporated this additional data to fine-tune our model and re-issue polynomial power profiles. The results from this second batch of validation experiments are presented in Section \ref{sec:batch_2}.

\begin{table}[]
\centering
\begin{tabular}{c|c|c|rrrr|c|c}
\hline
\multicolumn{1}{l|}{Batch} & Condition & Fig. & \multicolumn{1}{c}{$\phi_1$} & \multicolumn{1}{c}{$\phi_2$} & \multicolumn{1}{c}{$\phi_3$} & \multicolumn{1}{c|}{$\phi_4$} & \multicolumn{1}{c|}{MAPE (\%)} & \multicolumn{1}{c}{Max T}\\ \hline
1 & 775C/Fast & \ref{fig:control_1}       & 1 & 1.14E-01&-1.80E-03&7.93E-06& 7.00 & 781C\\
1 & 750C/Fast & \ref{fig:control_2}       & 1 & 9.53E-02&-1.47E-03&6.63E-06& 20.5 & 756C\\
2  & 775C/Fast    & \ref{fig:control_3}   & 1 & 9.80E-02&-1.40E-03&5.87E-06& 10.7 & 778C\\
2  & 750C/Fast    & \ref{fig:control_4}   & 1 & 9.43E-02&-1.37E-03&5.78E-06& 9.29 & 754C\\
2  & 775C/Slow     & \ref{fig:control_5}  & 1 & 6.45E-02&-6.61E-04&2.13E-06& 12.1 & 783C\\
2  & 750C/Slow     & \ref{fig:control_6}  & 1 & 6.42E-02&-7.39E-04&2.58E-06& 17.4 & 756C\\
\hline
\end{tabular}
\caption{Summary of experiments and power profile polynomial candidates.}
\label{tab:summary}
\end{table}

\subsection{Batch \#1} \label{sec:batch_1}
Seven FSP experiments were performed using simple linear power profiles (\textit{i.e.}, $P\left(t;\phi \right) = \phi_1 + \phi_2 t$) that had been hand-tuned. These initial experiments covered a range of desired set tool temperatures in the range of 700 to 800 degrees Celsius in a 316L stainless steel plate. Pursuant to the system ID task outlined in Section \ref{sec:system_id}, we tune Eq. \ref{eq:system} to best reproduce all seven unique experiments. Full details on this dataset are presented in Koch et al. \cite{koch2024neural}.

\begin{figure}[t]
        \centering
        \begin{overpic}[width=6in]{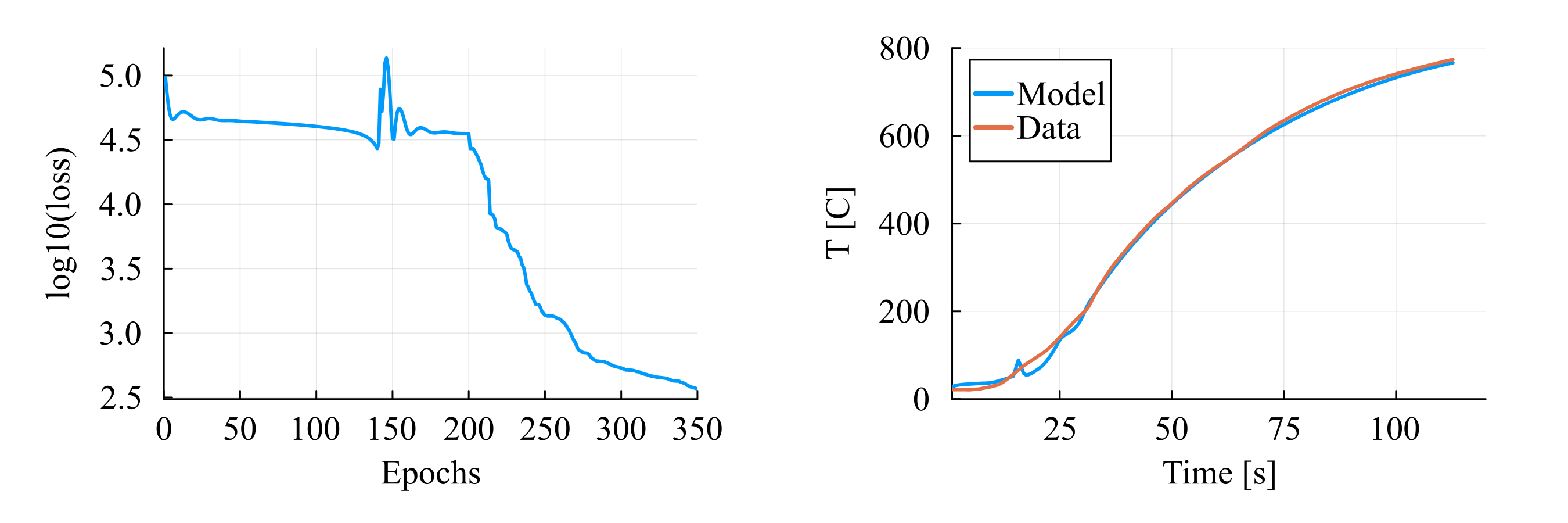}  
        \put(1,30){(a)}
        \put(50,30){(b)}
        \end{overpic}  
        \caption{Training loss history for the system ID task is shown in (a). An example trajectory reconstruction (given actual power input) is shown in (b). The Mean Absolute Error over the 7-run dataset is approximately 15 degrees Celsius.}
        \label{fig:training_1}
\end{figure}

\begin{figure}[t]
        \centering
        \begin{overpic}[width=4in]{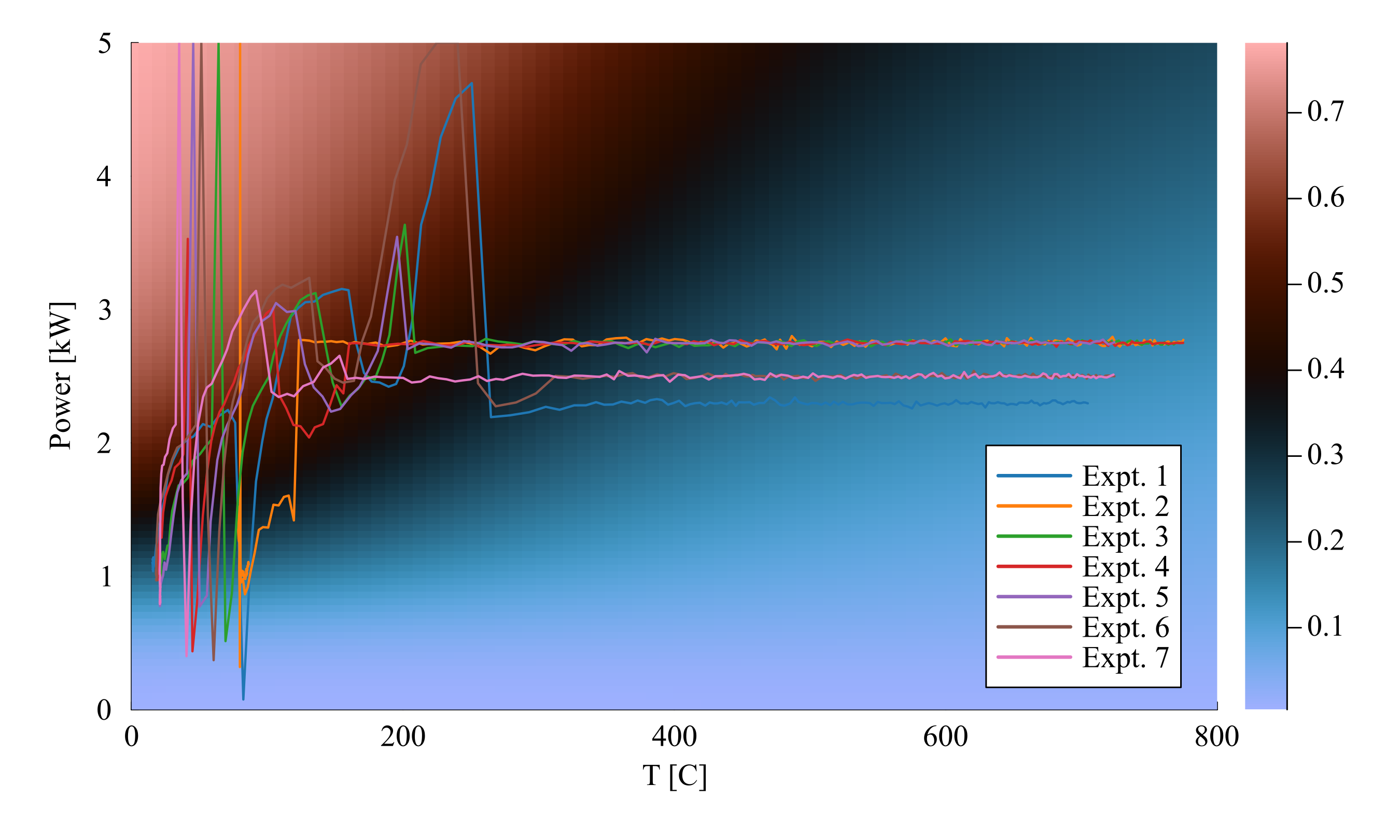}  
        \end{overpic}  
        \caption{The system dynamics model of Eq. \ref{eq:system} contains a neural network stand-in for the unknown conversion factor between input power and internal heating. This neural network accepts power and temperature and outputs a conversion efficiency, $\eta \in (0,1)$. The heatmap depicts the trained neural network response over this temperature-power space. To get a sense for how experiments behave in this space, overlaid are the seven original experimental time series in the same temperature-power coordinates.}
        \label{fig:heatmap_1}
\end{figure}

Figure \ref{fig:training_1} depicts the training history for the system ID model alongside an example trajectory reconstruction. The reconstruction error - in terms of Mean Absolute Error - was 15.6 degrees Celsius across all runs in the initial dataset. The trained model contains a tuned 2-dimensional response surface (a Neural Network) that maps tool temperature and power input to a power conversion efficiency $\eta$. Figure \ref{fig:heatmap_1} depicts this response surface with all seven experiments overlaid in the Temperature-Power space. 

The desired set temperatures for Batch \#1 experiments were 775 and 750 degrees Celsius. These set temperatures did not appear in the training dataset. The loss scaling factors corresponding to Eq. \ref{eq:control} were $\lambda_{1\rightarrow 4} = (1, 1\cdot10^3,1\cdot10^6,1\cdot10^3)$. The model-informed power profiles are summarized in Table \ref{tab:summary}. 

\begin{figure}[h]
\centering
\begin{minipage}{.5\textwidth}
  \centering
  \includegraphics[width=1\linewidth]{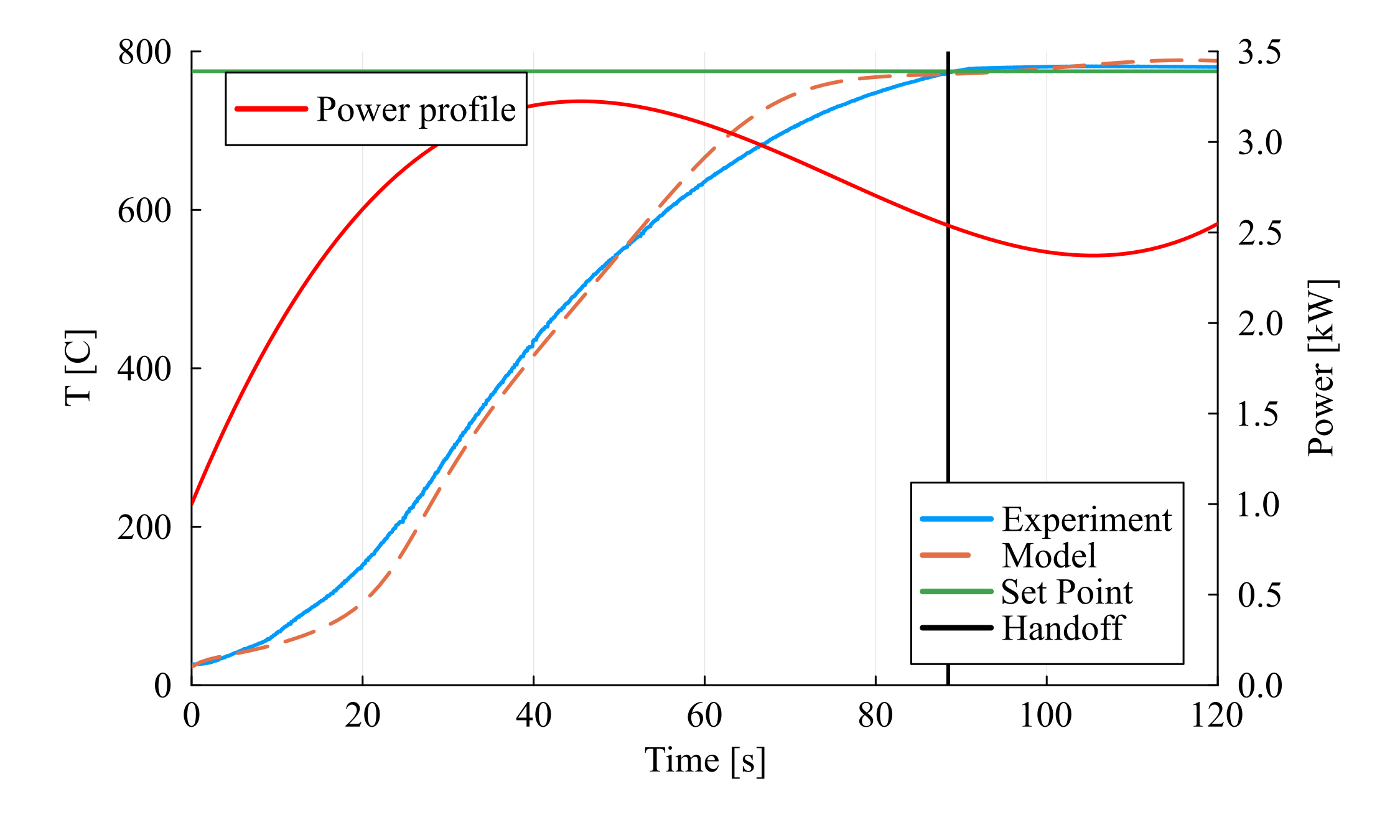}
  \captionof{figure}{Batch \#1. 775C set point.}
  \label{fig:control_1}
\end{minipage}%
\begin{minipage}{.5\textwidth}
  \centering
  \includegraphics[width=1\linewidth]{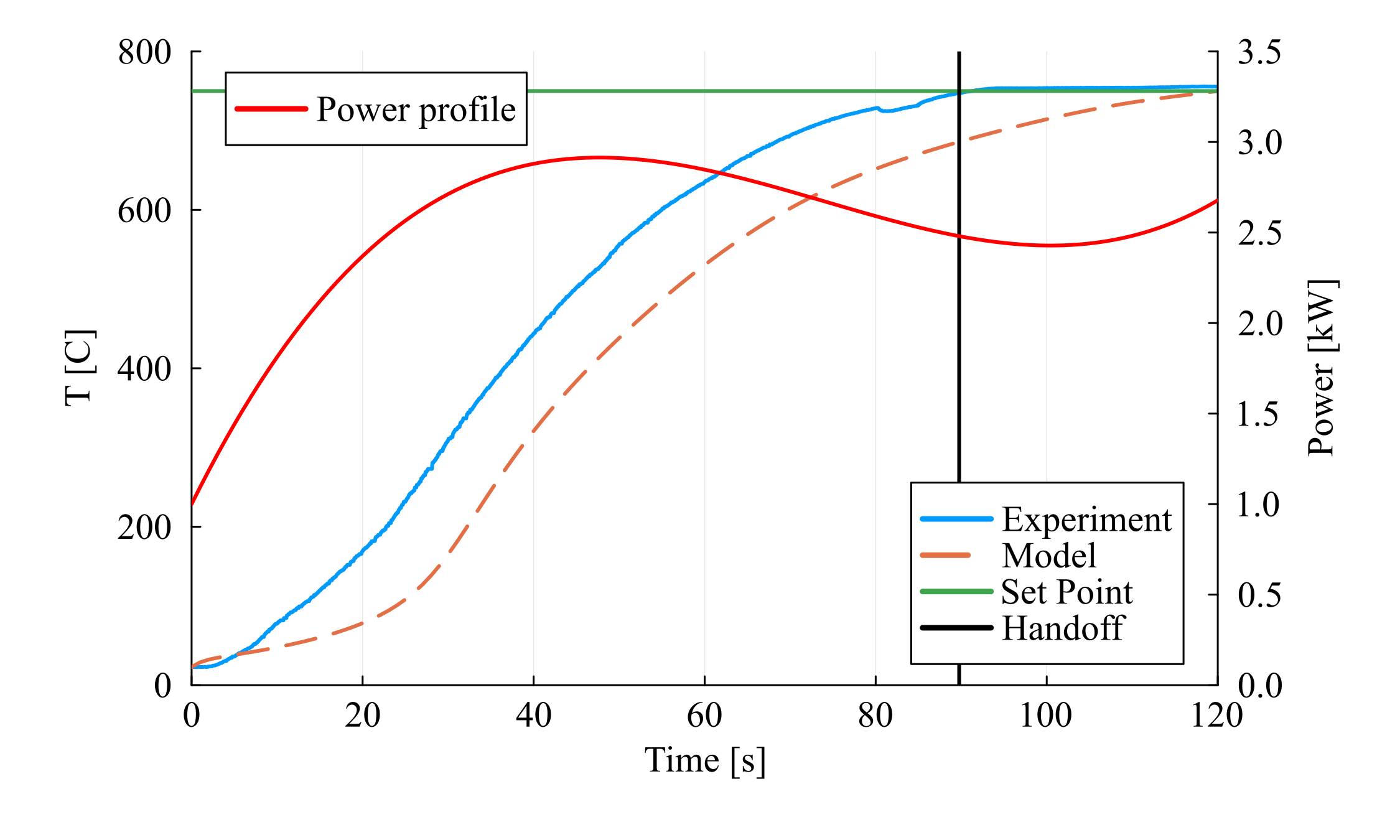}
  \captionof{figure}{Batch \#1. 750C set point.}
  \label{fig:control_2}
\end{minipage}
\end{figure}

Figures \ref{fig:control_1}-\ref{fig:control_2} show the comparison between the experimentally-recorded tool temperatures via a thermocouple located in the FSP tool shoulder, the model prediction, and the desired set points. In these figures, the time series data for the model and experiment are co-located in time where the control power profile begins. The black vertical bar denotes the location in time where the FSP machine transitions from power control to temperature control; \textit{i.e.}, the ``handoff''.

The Mean Absolute Percent Errors (MAPE) between the experimental temperature data and the model predictions from the tool \textit{first contact} (time $t=0$) to the ``handoff'' to the traversal stage of the weld (denoted by the black vertical bars) is summarized in Table \ref{tab:summary}. Note that while the overall system behavior and shape of the temperature time series are consistent across experiment and the model, the model-experiment mismatch in the initial ~20 seconds of the plunge amplify error metrics of this type. End-of-plunge temperatures for both experiments were within 1\% of the desired setpoint.

\subsection{Batch \#2} \label{sec:batch_2}
The experiments from Batch \#1 were added to the initial training set of 7 experiments for a total of 9 unique experiments. The system ID and control optimizations were re-run with the new data. For Batch \#2, we sought to reach the same 750 and 775 degrees Celsius setpoints through ``fast'' and ``slow'' power profiles; that is, promoting more or less aggressive heating to reach the desired end temperature. To do so, we adjusted our loss hyperparameters as:
\begin{equation} \label{eq:hyperparams}
\begin{split}
\lambda_{1\rightarrow 4, \text{fast}} &= (1, 1\cdot10^3,1\cdot10^6,1\cdot10^3) \\
\lambda_{1\rightarrow 4, \text{slow}} &= (1, 1\cdot10^3,1\cdot10^7,1\cdot10^3) .
\end{split}
\end{equation}
Here, we increased the smoothing penalty scaling by an order of magnitude between the ``fast'' and ``slow'' cases.

\begin{figure}[h]
\centering
\begin{minipage}{.5\textwidth}
  \centering
  \includegraphics[width=1\linewidth]{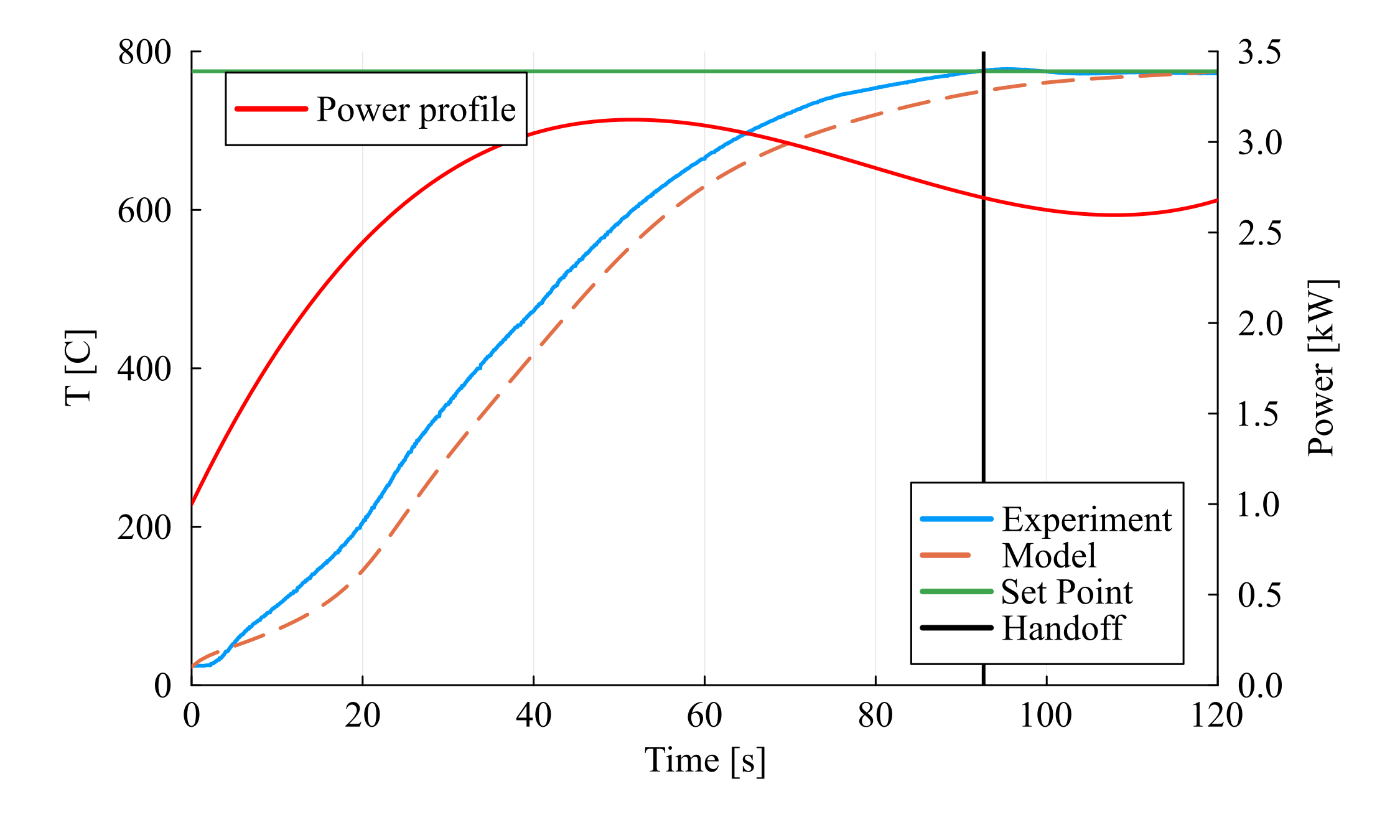}
  \captionof{figure}{Batch \#2. 775C set point (fast).}
  \label{fig:control_3}
\end{minipage}%
\begin{minipage}{.5\textwidth}
  \centering
  \includegraphics[width=1\linewidth]{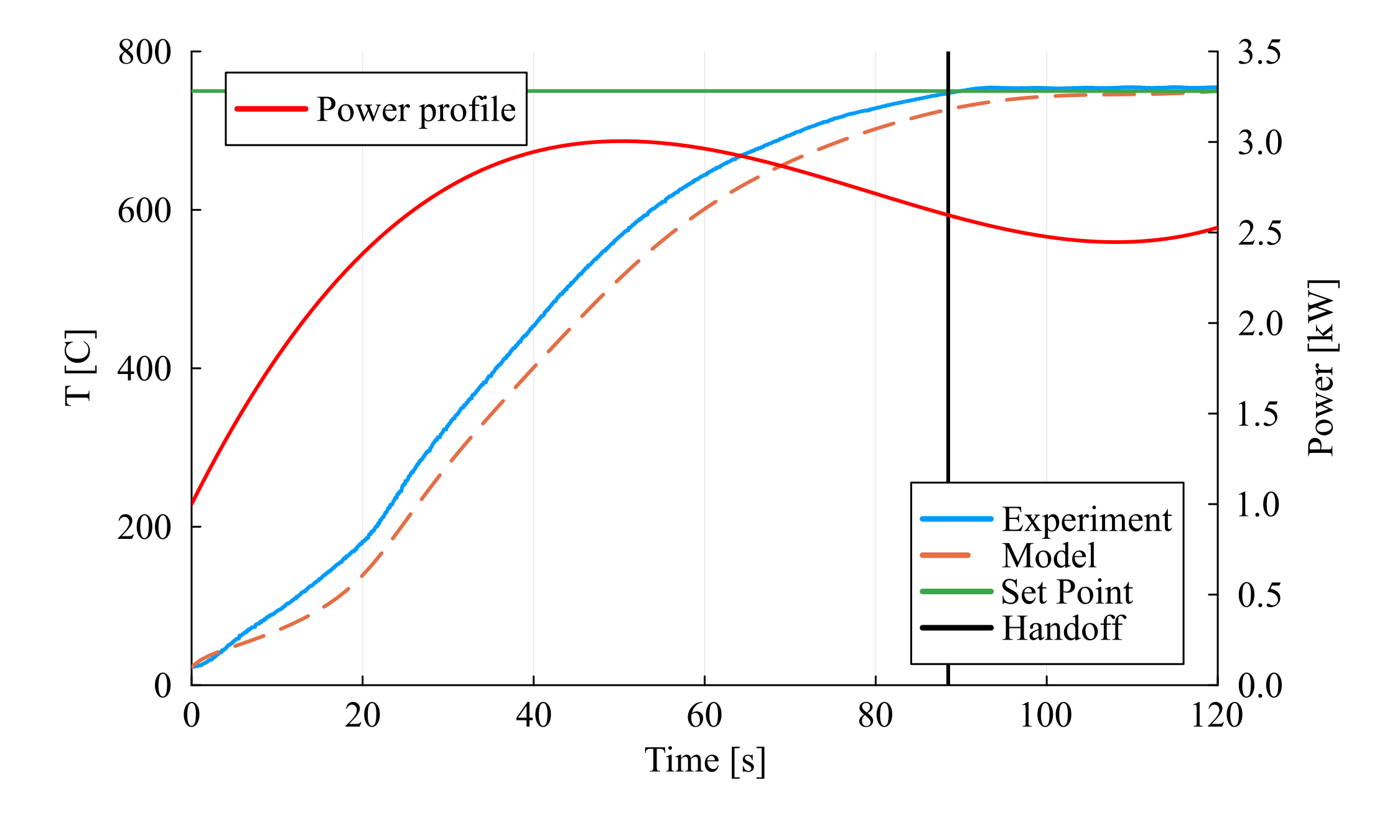}
  \captionof{figure}{Batch \#2. 750C set point (fast).}
  \label{fig:control_4}
\end{minipage}
\end{figure}

\begin{figure}[h]
\centering
\begin{minipage}{.5\textwidth}
  \centering
  \includegraphics[width=1\linewidth]{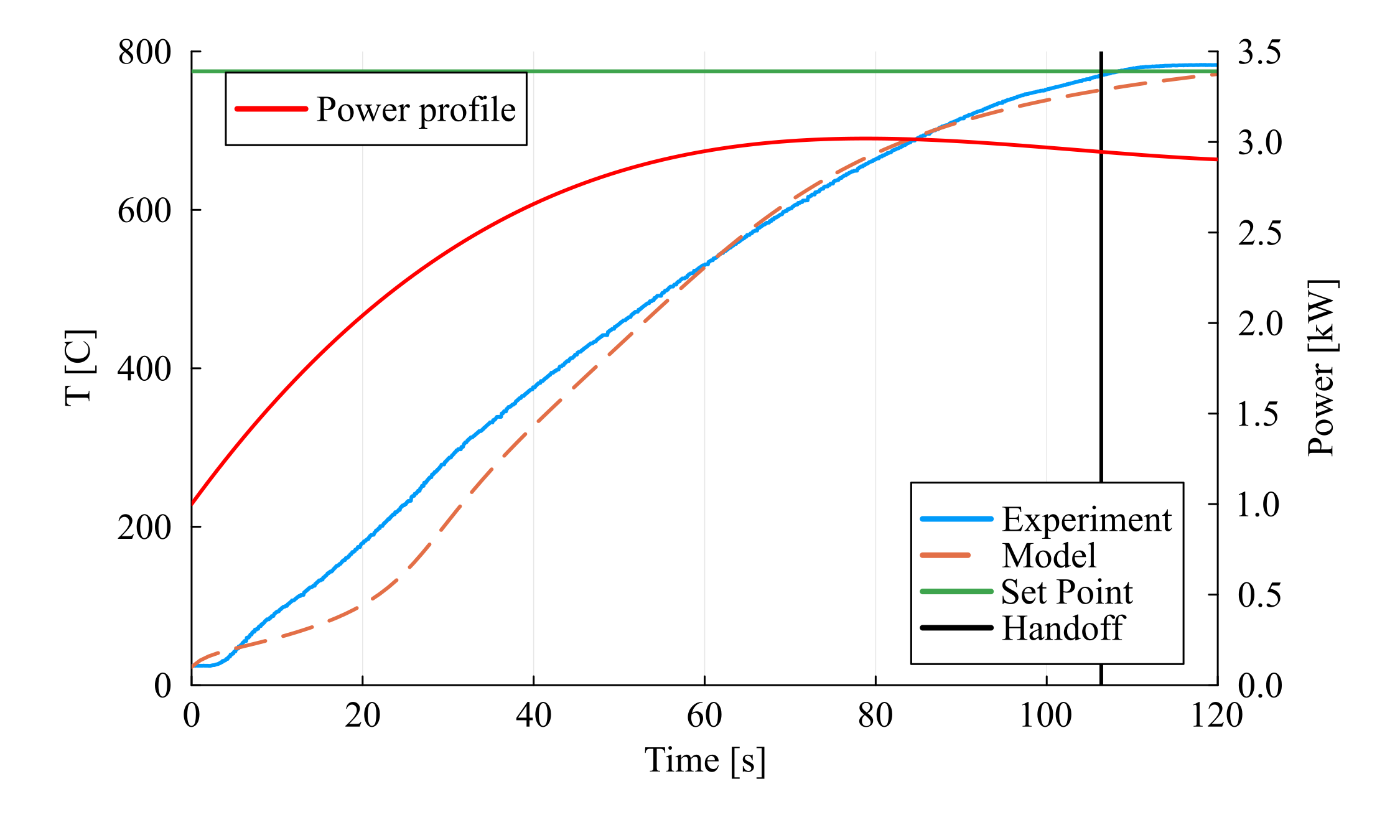}
  \captionof{figure}{Batch \#2. 775C set point (slow).}
  \label{fig:control_5}
\end{minipage}%
\begin{minipage}{.5\textwidth}
  \centering
  \includegraphics[width=1\linewidth]{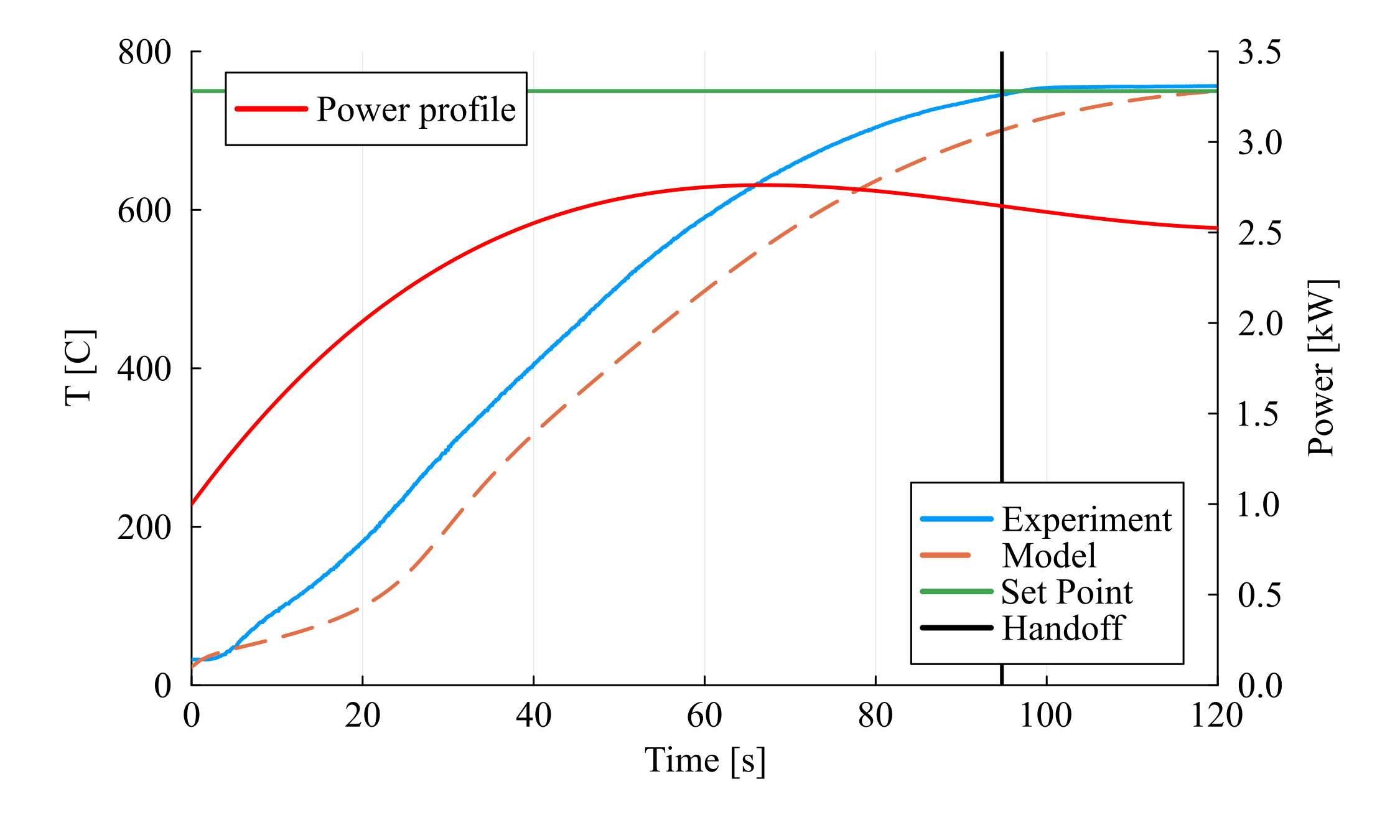}
  \captionof{figure}{Batch \#2. 750C set point (slow).}
  \label{fig:control_6}
\end{minipage}
\end{figure}

The results of the second batch of experiments is summarized in Table \ref{tab:summary} and visually depicted in Figs. \ref{fig:control_3}-\ref{fig:control_6}. Across this batch, the MAPE for the duration of the plunge sequence was consistent at around 12.5\%. Note that this error metric is highly sensitive to the initial contact phase of the experiment. Although experiment and model temperatures qualitatively match very well, time delays in the model temperature are the dominant contributor to this error metric. The final handoff temperatures were at or below 1\% of the desired setpoint for all experiments in this batch. 

\section{Discussion and Conclusions}
This work validated the capability of data-enabled, simplified system dynamics models to perform real-world control tasks for friction-stir processing. Where traditional methods for modeling and control of such systems require extensive first-principles modeling, simulation, and experimental validation campaigns, our approach has simplified and accelerated this pipeline while remaining interpretable and traceable to first-principles physics. The core methodology involves neural lumped-parameter differential equations, whereby we construct a model that accounts for energy pathways in the FSP dwelling stage that are automatically tuned to match experimental observables (temperature time series) by virtue of the differentiability of the model. The methodology comprises two stages: (i) system identification, where the model is tuned to fit the observational data, and (ii) model predictive control, where the model -- with frozen weights -- is used to compute optimal control policies. This work focused specifically on open-loop control, but we assert that closed-loop control is a natural extension to this work, provided necessary experimental infrastructure exists.

This work opens up several avenues for high-impact future research. First, exploring closed-loop feedback control can greatly enhance controller performance (\textit{e.g.}, adaptability and error correction). Second, as presented, our experimental campaign was specific to a particular stainless steel alloy, namely 316L. The flexibility of data-driven methods allows for greater model parameterization and transfer learning, which can enable generalization across different materials and processing conditions. Similarly, integration with real-time data acquisition and compute can enable real-time analytics and quality control, where data streams are used both for control as well as persistent monitoring (\textit{e.g.}, defect detection and avoidance). 

\section*{Acknowledgements}

The research described in this paper is part of the Materials Characterization, Prediction, and Control agile investment at Pacific Northwest National Laboratory. It was conducted under the Laboratory Directed Research and Development Program at PNNL, a multiprogram national laboratory operated by Battelle for the U.S. Department of Energy under contract DE-AC05-76RL01830.

\bibliographystyle{elsarticle-num} 
\bibliography{fsw}

\newpage
\appendix

\end{document}